\documentclass[runningheads]{svmult}

\usepackage{makeidx}   

\usepackage{graphicx}  

\usepackage{subeqnar}  

\usepackage{multicol}  

\usepackage{cropmark} 

\usepackage{physprbb}  

\makeindex             

\begin{document}

\title*{New results on the temporal structure of GRBs}

\titlerunning{New results on the temporal structure of GRBs}


%

\author{E. Nakar \& T. Piran}

\institute{Racah Institute, Hebrew University, Jerusalem 91904, Israel}

\maketitle              

\begin{abstract}
  We analyze the temporal structure of long (\( T_{90}>2sec \)) and
  short (\( T_{90}<2sec \)) BATSE bursts.  We find that: (i) In many
  short bursts \( \delta t_{min}/T\ll 1 \) (where \( \delta t_{min} \)
  is the shortest pulse). This indicates that short bursts arise, like
  long ones, in internal shocks.  (ii) In long bursts there is an
  excess of long intervals between pulses (relative to a lognormal
  distribution). This excess can be explained by the existence of
  \emph{quiescent times}, long periods with no signal above the
  background that arise, most likely, from periods with no source
  activity.  The lognormal distribution of the intervals (excluding
  the \emph{quiescent times}) is similar and correlated with the
  distribution of the pulses width, in agreement with the predictions
  of the internal shock model.
\end{abstract}

\section*{Introduction}

The variability of GRBs provided the main evidence for the  internal-external shocks scenario.
External shocks cannot produce efficiently such variability \cite{1}.
Internal shocks can produce such temporal structure provided that there are two time scale within
the  ``inner engine" - a short time scale that produces the variability and a long time
scale that determines the duration of the burst. 
So far variability was shown only for long bursts.  It
is an open question whether short bursts arise in internal shocks as well.
Using a new algorithm \cite{2} we study their variability.
We also present some new results on the temporal structure of long bursts. 
Our results provide further support for the internal shocks scenario and show that three different time scales operate within the ``inner engine". 

\section*{Variability of short bursts}

We analyze the distribution of  \( \delta t_{min}/T \) (where $\delta t_{min}$ is the duration of the
shortest pulse in a burst, and  $T$ is the total duration of the burst)
in a sample of the brightest 33 short bursts
(peak flux in 64ms$>$4.37\( ph/(sec\cdot cm^{2}) \)) with a good TTE data coverage
(for BATSE data types review \cite{3}). The TTE data is binned into 2msec
time bins. We compare the results to a sample of 34 long bursts with the same
peak flux, using the 64ms concatenate data to which we have  added a Poisson noise 
so that the signal to noise ratio of both samples would be similar.
We call this later sample the  \emph{'noisy long'} set.
\begin{figure}
{\par\centering \resizebox*{0.9\columnwidth}{0.2\textheight}{\includegraphics{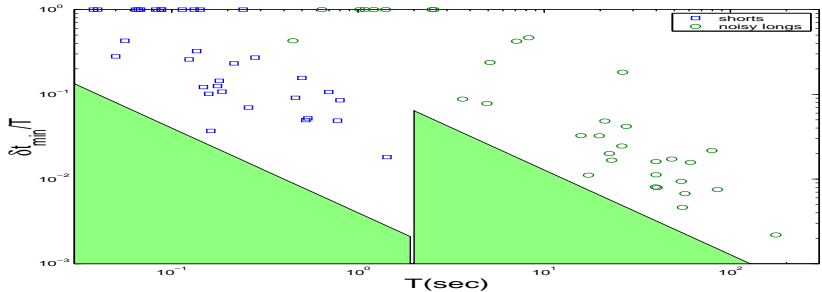}} \par}

\caption{\label{dt/T in shorts}{\small The shortest pulse-\protect\( \delta t_{min}\protect \)
represented as a function of the total duration of the burst. The shaded areas
are excluded because of the data resolution (4ms for shorts and 128ms for noisy
longs) or the data definition (\protect\( \delta t_{min}\leq T\protect
\)). Notice that: (i) $\delta t_{min}$ and $T$ are not correlated. (ii) Some of
the results are approaching the resolution limit. (iii) In some of the
bursts $\delta t_{min}=T$. }\small }
\end{figure}

Fig~\ref{dt/T in shorts} depicts  \( \delta t_{min}/T \)
in both data sets: \emph{'short'} and \emph{'noisy long'}. In the \emph{'short'}
set the median of \( \delta t_{min}/T \) is 0.25.  35\% of bursts have
\( \delta t_{min}/T<0.1 \) and 35\% of the bursts show a smooth structure
(\( \delta t_{min}/T=1 \)). This result could mislead us to the conclusion
that a significant fraction of the short bursts have a smooth time profile.
But a comparison with  the \emph{'noisy long'} results show that also in this group more
than 20\% of the bursts are single pulsed, while there were no such bursts in
the original (without the added noise) \emph{'long'} set. 

We conclude that short bursts are variable and hence are most likely produced in internal shocks.
While the observed variability is not as large as seen in long bursts one has to remember that when
studying variability of short bursts we are approaching the instrumental limitations both in terms
of the time scales and of the signal to noise ratio. It
is possible that  10\%-20\% of the short bursts are produced by external shocks.

\section*{The pulses' width  and the intervals between pulses}

According to the internal shocks model \cite{4} the source ejects relativistic
shells with different velocities and shocks arise when faster shells catch slower ones.
We show in \cite{5} that  both the pulses` width
\( \delta t \) and the intervals between pulses
\( \Delta t \) are proportional to the same parameter -
the separation between two following shells, namely the variability time scale
of the ``inner engine". Therefore both distributions
should be similar. Moreover, any interval should be correlated to the width of its neighboring pulses.

We have applied our algorithm to a sample of the 68 brightest long bursts in
BATSE 4B catalog (peak flux in 64ms$>$10.19\( ph/(sec\cdot cm^{2}) \)) . This
resulted in 1330 pulses (1262 intervals). Our null
hypothesis was that both $\delta t$ and $\Delta t$, have lognormal distributions.
The \( \chi ^{2} \) test gives
a probability of \( 0.52 \) that the pulses width were taken from a lognormal
distribution with \( \mu =0.065\pm 0.04 \) (\( \overline{\delta t} \approx 1sec \)) and \( \sigma =0.77\pm 0.03 \)
(\( 1\sigma  \) corresponds to $\delta 
t$ between 0.5 and 2.3sec).

The  \( \Delta t \) distribution shows, however,   an excess
of long intervals relative to~a~lognormal distribution.
The \( \chi ^{2} \)  probability for a lognormal distribution is
\textbf{\( 1.2\cdot10^{-10} \)}.
McBreen\cite{6,7} and Li \& Fenimore\cite{8} suggest that
this deviation is due to the limited resolution (64ms). However, fitting
the intervals above the median with a half Gaussian fails. The inconsistency is not due to the resolution.

Many of the long intervals are dominated by a quiescent time: periods within the burst with no observable counts above the background
noise. When excluding \emph{all} the intervals that
contained a quiescent time  the \( \chi ^{2} \)  probability
that the data is lognormal
is \textbf{\( 0.27 \)}, with \( \mu =0.257\pm 0.051 \) (\( \overline{\Delta t} \approx 1.3sec \))
and \( \sigma =0.90\pm 0.04 \) (\( 1\sigma  \) corresponds to  $\Delta 
t$ between 0.53 and 3.1sec).

The similarity between the parameters of both distributions is remarkable. Moreover,
we find, as predicted by the internal shocks model, a linear correlation,
\( r \), between intervals and the following pulses.
The average \( r \) is 0.48, showing a strong correlation.

\section*{Conclusions}
For most short bursts \( \delta t_{min}/T\ll 1 \).
This suggests that these bursts are produced
by internal shocks.
If, later, the ejecta encounters a surrounding ISM then we expect it to produce an
external shock and emit an afterglow.
For some (30\% of our sample) short
bursts \( \delta t_{min}\approx T \). However, a comparison with the \emph{'noisy
long'} set, shows that this feature could very well be due to the noise.
We cannot rule out the possibility that 10\%-20\% of the short bursts are produced
by external shocks or by a single internal collision.

The distribution of interval between pulses shows an excess of long intervals relative to
a lognormal
distribution. After removing intervals that include quiescent times the distribution
is consistent with a lognormal distribution with comparable parameters to the
pulse width distribution.
This result suggests that the \( \Delta t \)
distribution is made from the sum of two different distributions: A lognormal
distribution that  is also compatible with the \( \delta t \) distribution and
the quiescent times distribution. As \( \Delta t\)
reflects the central engine behavior, this suggests that there are two different
mechanisms operating within the source. A short time scale  mechanism,
with a lognormal distribution and a longer time
scale mechanism that turns the  central engine on
and off and
is responsible for the quiescent times. Results that support this
suggestion were obtained by Ramirez \& Merloni \cite{9} .
The correlation
between the interval and the following pulse
confirms this suggestion and is in an excellent
agreement with the internal shocks model.

\end{document}